**Title: A windowed local fdr estimator providing higher resolution and robust thresholds**

Rishi L. Khan[1,2], Rajanikanth Vadigepalli[1], Guang Gao[2] and James S. Schwaber[1]

[1]Department of Pathology, Anatomy and Cell Biology, Thomas Jefferson University, Philadelphia, Pennsylvania 19107, USA

[2]Department of Electrical Engineering, University of Delaware, Newark, Delaware 19716, USA

Corresponding author:
James Schwaber
Director, Daniel Baugh Institute for Functional Genomics/Computational Biology
Department of Pathology, Anatomy and Cell Biology
Thomas Jefferson University
1020 Locust Street Room 381
Philadelphia, PA 19107
215-503-7823
FAX: 215-503-2636
Email: james.schwaber@mail.dbi.tju.edu

**ABSTRACT**

**Motivation**. In microarray analysis, special consideration must be given to the issues of multiple statistical tests and typically p-values are adjusted to control family-wise error rate (FWER) or false discovery rate (FDR). FDR metrics have been suggested for controlling false positives, however, genes with p-values close to the threshold typically have a higher chance of being false positives than genes with very low p-values. The local FDR (*fdr*) metric gives the number of false positives in the vicinity of a test statistic. We propose a new *fdr* estimator that uses windows instead of binsand define heuristics that use the fluctuations in the estimator to determine robust thresholds for classifying differential expression.

**Results**: Our *fdr* approach estimates the false discovery rate within a window of p-values. We present heuristics that derive robust *fdr* thresholds such that a significant change in the *fdr* threshold yields a small change in the number of rejected hypotheses. We compare these thresholds with thresholds from other approaches using two simulated datasets and one cancer microarray dataset. In the latter, our estimator finds two robust thresholds. Since our *fdr* estimator is an extension of the FDR metric, it can be used with many FDR estimation methods.

**Availability**: An R function implementing the proposed estimator is available at http://www.dbi.tju.edu/dbi/tools/fdr

**Contact**: james.schwaber@jefferson.edu

**Supplementary Information:** Supplementary figures and code are available at http://www.dbi.tju.edu/dbi/tools/fdr

**BACKGROUND**

We present a local false discovery rate (*fdr*) estimator that provides an improved metric of differentially expressed genes, permitting an estimate of false discovery rate in a local window of probabilities. DNA microarray data analysis requires testing of thousands of hypotheses relating to differential gene expression. However, statistical results must be analyzed using multiple test corrections to take into account false positive error rates. The first methods for analyzing multiple tests controlled the family wise error rate (FWER), which is the probability of incorrectly rejecting at least one null hypothesis (type I error) in the family of all rejected hypothesis (Dudoit, et al., 2002; Hsu, 1996) assuming the tests are independent. Westfall and Young (Westfall and Young, 1993) proposed a method of resampling the data to estimate the dependence structure while controlling the FWER. However, all of these methods have a high false negative, or "miss rate" because they control the probability of making a single error (Taylor, et al., 2005).

The false discovery rate (FDR) is the proportion of incorrectly rejected null hypotheses in the family of all rejected hypothesis. Benjamini and Hochberg developed a method to provide strong control of the false discovery rate under independence and suggested that this metric may be more suitable than FWER control for large screening studies (Benjamini and Hochberg, 1995). There are other methods of estimating and controlling the false discovery rate such as resampling based methods to account for correlation in the test statistics (Reiner, et al., 2003; Tusher, et al., 2001; Yekutieli and Benjamini, 1999), FDR control under arbitrary dependence (Benjamini and Yekutieli, 2001), accounting for the estimated number of false null hypothesis (Storey, 2003), and rank invariant resampling (Jain, et al., 2005).

**Motivation for local false discovery rate (fdr)**

The FDR metric controls the global false discovery rate of all rejected hypotheses. However, the hypotheses with higher p-values have higher rates of false positives than hypotheses with very low p-values. To illustrate this, we present a simple scenario (adapted from the comments to (Ge, et al., 2003)).

Suppose there are 8000 genes and 500 of them are differentially expressed and up regulated exactly 8-fold. However, due to measurement noise, the $\log_2$-ratio measurements of the differential expression come from the normal distribution $N(\mu=3, \sigma=0.5)$ [note that $\log_2(3)=8$], and the measurements from the other genes come from the normal distribution $N(\mu=0, \sigma=0.5)$. We measure the expression of each gene 5 times and take the two-sided one sample Student's T-test with the null hypothesis that $\mu=0$. Figure 1a shows the sorted p-values as a function of the number genes classified as differentially expressed at each p-value.

FDR control (Benjamini and Hochberg, 1995) at $\alpha=0.2$ will yield a p-value threshold that classifies 625 genes as differentially expressed: all 500 differentially expressed genes are classified correctly and 125 genes that are not differentially expressed are classified as differentially expressed (125/625 = 20% FDR). If we sort the genes by p-value, the first 500 genes contain 8 genes that are not actually differentially expressed (FDR=1.6% in window). The next 50 genes have 8 that are differentially expressed (FDR=16% in window). The next 75 genes have no genes that are differentially expressed (FDR=100% in window). If the false discovery rate in a window could be estimated, then it would be possible to avoid accepting excessive false positives after most of the genes have been correctly classified. From Figure 1b, rejecting the null hypothesis of genes beyond the first 500 yields unsatisfactory results. Figures for other sample sizes are presented in the supplementary data (simulated.bimodal.pdf).

The local FDR metric was first proposed by (Efron, et al., 2001) as $p_0 f_0(Z)/f(Z)$ where $p_0$ is the number of genes differentially expressed, $f_0$ is the distribution of z statistics of the genes that are not differentially expressed, $f$ is the distribution of z statistics of all genes, and $Z$ is the interval at which the local false discovery rate is measured (e.g. $1.9<Z<2.1$). However, these posterior probabilities ($p_0$, $f_0$, and $f$) are not known and must be estimated. There are many different ways of estimating local FDR including using parametric mixture models (Liao, et al., 2004; Pan, 2002; Pounds and Morris, 2003), empirical distributions (Scheid and Spang, 2004), and nonparametric models based on LOESS (Pounds and Cheng, 2004). (Broberg, 2005) has a review of many of the pertinent methods. Many of these methods use binning and smoothing to obtain an estimate that produces a monotonic function. However, we propose a method that can increase resolution by using a sliding window and does not smooth the estimate but uses fluctuations in the estimate to propose robust thresholds.

**A local fdr estimator**

Let:

$f_0$ = distribution of p-values of genes that are not differentially expressed

$f_1$ = distribution of p-values of genes that are differentially expressed

$f$ = distribution of p-values of all genes

$\pi_0$ = fraction of genes that are differentially expressed

$f(p) = \pi_0 * f_0(p) + (1-\pi_0) * f_1(p)$

By definition, $f_0$ is the uniform distribution. $f$ is the empirical distribution of actual p-values obtained from a microarray experiment. $\pi_0$ and $f_1$ are unknown and must be estimated. There are many ways to estimate $\pi_0$ (Broberg, 2005), but we choose to use a simple method here and leave this to other literature.

We note that the distribution of p-values of differentially expressed genes approaches 0 for intervals close to 1. If we choose a range of p: [λ,1] such that $f_1([\lambda,1]) \approx 0$, then $f_0([\lambda,1]) = 1-\lambda$ and

$$f([\lambda,1]) = \#\{p_i > \lambda\}/N \approx \pi_0(1-\lambda)$$ where $0<\lambda<1$ and N is the total number of genes.

Therefore:

$$\hat{\pi}_0(\lambda) = \frac{\#\{p_i > \lambda\}}{(1-\lambda)N}$$

The choice of λ such that $f_1([\lambda,1]) \approx 0$ is one of much debate and we take the simplistic approach of choosing the median of the distribution $\pi_0(\lambda)$ for $0.25 < \lambda < 0.75$. This method tends to work well in practice. A larger discussion can be found in (Storey and Tibshirani, 2003).

We define local false discovery rate within an interval of p-values $[p_a, p_b]$ as:

$$fdr([p_a, p_b]) = \frac{\pi_0 f_0([p_a, p_b])}{f([p_a, p_b])}$$

Let us order all p-values obtained for microarray anaylsis $0 < p_1 \le p_2 \le \ldots \le p_n \le 1$. Let *ws* denote a window size. Since $f_0$ is uniform over [0,1]:

$$fdr([p_{i-ws}, p_i]) = \frac{\pi_0 f_0([p_{i-ws}, p_i])}{f([p_{i-ws}, p_i])} = \frac{\pi_0(p_i - p_{i-ws})}{ws/N} = \frac{N\pi_0(p_i - p_{i-ws})}{ws}$$

This can be written as a modification of the FDR control method:

If $FDR(i) = \frac{N\pi_0 p_i}{i}$, then

$$fdr(i, ws) = \pi_0 * \frac{FDR(i) * i - FDR(i-ws) * (i-ws)}{ws}$$

We assign *fdr* of a window to the last gene in the window (i.e. *fdr*(i) = *fdr*([$p_{i-ws}$,$p_i$])). Other approaches may assign the *fdr* of a window to genes at the beginning or middle of the window. However, our approach will be more conservative.

The *fdr* method is an extension of the FDR method, and therefore can work with other methods that control FDR such as the FDR control method that uses the dependence structure of the data (Benjamini and Yekutieli, 2001) or the positive FDR (pFDR) method (Storey, 2003). Note that Storey's pFDR method often yields large q-values for the genes with the lowest p-values resulting in negative *fdr* estimates. However, the adjusted q-values are montonically increasing and are appropriate to be used with the *fdr* method.

We will show in the Case Studies section that the local *fdr* metric is a much more robust metric than FDR for choosing a p-value threshold. That is, the number of rejected hypotheses does not change significantly when choosing a p-value threshold at local *fdr* = 20% or local *fdr* = 50%. This is not the case for FDR.

**Heuristics for assigning a p-value threshold given a local fdr threshold**

The local *fdr* estimator we have defined produces a non-monotonic result. We provide heuristics for choosing a p-value threshold given a local *fdr* threshold. The case studies focus on microarray gene expression data, but the local *fdr* approach is applicable to any data types involving multiple hypotheses testing.

Let the nth intersection of the *fdr* estimate and the *fdr* threshold be $i_n$. Let the number of genes classified as differentially expressed at this intersection be $g_n$. Let $p_n$ be the p-value of the gene $g_n$. Let n=1.

1) If the *fdr* estimate crosses the *fdr* threshold only once, use the p-value of this gene for the p-value threshold. Go to step 5. Otherwise, the *fdr* estimate crosses the *fdr* threshold more than once. Go to step 2.

2) If $g_{n+1} - g_n < ws$, then go to step 4. Otherwise go to step 3.

3) Calculate the local *fdr* in the window between $i_n$ and $i_{n+1}$ as described above. If this local *fdr* is less than the *fdr* threshold then go to step 4. If this local *fdr* is greater than the *fdr* threshold then use $i_n$ and classify $g_n$ genes as differentially expressed. Choosing the $i_{n+1}$ would yield more false positives than acceptable. Go to step 5.

4) Increment n. If there are no more intersections, go to step 5. Otherwise go to step 2.

5) Use $p_n$ as the p-value threshold.

We can find $p_n$ for all *fdr* thresholds between 0 and 1. In the results section, we will apply these heuristics to simulated data. Figure 3c shows *fdr* threshold value and FDR threshold value versus the number of genes classified as differentially expressed. Note that number of genes classified as differentially expressed is more robust to the local *fdr* than the FDR.

We have presented a method for estimating the local false discovery rate (*fdr*) in a window. This metric should be used in conjunction with the standard FDR metric. In the example above, the *fdr* would choose a threshold somewhere in the grey area of Figure 1a. The *fdr* metric gives an upper bound for useful FDR threshold values and if the *fdr* threshold given is controlling the FDR at an unacceptable $\alpha$ level, it should not be used. Typically in microarray analysis we can tolerate a range of FDR values from 0 to 20%.

The FDR metric does not give any robust thresholds within this range while the local *fdr* ordinarily does.

**CASE STUDIES**

We will use three datasets to compare these methods: two simulated datasets (small and large changes in gene expression) and one microarray study on cancer patients (Golub dataset (Golub, et al., 1999)). The Golub dataset is widely used as a benchmark in multiple hypothesis testing algorithms and provides a basis for comparison with previous literature.

**Simulated data**

The simulated gene expression data is based on typical characteristics of large-scale microarray data:

- A1. Most genes do not exhibit differential expression.
- A2. The differential expression of any particular gene across several replicate experiments is log-normal.
- A3. The standard deviation of differential gene expression across several replicate experiments is log-normal.

The true gene expression model is derived from microarray characteristic A1. A truth model T contains a gene expression dataset of N genes for an experimental sample TE and a control sample TC. The nominal control sample models this as:

The control sample TC is generated as follows:

TC = {$g_j$ : $\log_2(g_j)$= minE + rand\_j*(maxE - minE) , j=1…N, rand\_j $\in$ $U$(0, 1)}

where minE and maxE are the minimum and maximum expression values (in $\log_2$ space). We have chosen to make the control expression uniform in $\log_2$ space across the spectrum of possible gene expression values. We do this without loss of generality because the test statistic will compare a gene measurement in the control sample only to its respective measurement in the experimental sample.

The nominal experimental sample TE is generated as: $\log_2(TE) = \log_2(TC) \bullet TGE$

The vector TGE specifies the differential gene expression. Most of the elements of TGE are 0 and correspond to genes that are not differentially expressed. We use a mixture model where the differentially expressed genes follow a uniform distribution until a cutoff and then the density tapers down to zero at the maximum expression as shown in Figure 2a.

**Noise Model**

The noise-corrupted gene expression model is derived from microarray characteristics A2 and A3. The control sample $C_i$ and experimental sample $E_i$ are given as follows:

$C_i = \{g_{ij} : \log_2(g_{ij}) = \log_2(TC_{ij}) + \log_2(noise_{ij}), j=1\ldots N, noise_{ij} \in N(0, SDB_j)\}$

$E_i = \{g_{ij} : \log_2(g_{ij}) = \log_2(TE_{ij}) + \log_2(noise_{ij}), j=1\ldots N, noise_{ij} \in N(0, SDB_j)\}$

Where $SDB_j$ is the standard deviation of the additive noise introduced by biological variability and procedural variability of the jth gene (e.g. labeling, hybridization, etc). As the data considered in this study does not contain replicate slides of each biological sample, we can only determine a lumped noise distribution, which we model by:

$\log_2(SDB_j) \in N(\mu,\sigma)$

where $\mu$ and $\sigma$ are estimated from microarray data. Figure 2b shows that this model fits microarray data well.

We will generate a gene expression dataset containing R replicates of noise corrupted experiment-control pairs. The truth model remains constant across all R replicates.

**Parameter Identification of the Noise Model and Simulation Framework**

We test the sensitivity and false discovery rate of the Students t-test with various p-value thresholding methods in the context of physiological responses in the brain. We choose to study 8,000 genes (because that is the current capacity of our arrays) and expect on the order of 500 to be differentially expressed. We have used microarrays to estimate the noise parameters as $\mu=-2.8$ and $\sigma=0.6$. Figure 2b shows a histogram of the standard deviations of differential gene expression measurements for each gene on our array over 6 experimental replicates (12 data points). The solid line is the log normal distribution with $\mu=-2.8$ and $\sigma=0.6$. We measure the effect of sample size (R) between 2 and 14 data points (1 and 7 biological sample pairs).

**Analysis methods**

In this study, we restrict ourselves to the Student's T-test as the test statistic for calculating p-values. For each gene containing R replicates, we perform the Student's T-test and obtain a p-value. All genes with Student's T-test p-value less than threshold $\alpha$ are classified as differentially expressed (or positive). All other genes are classified as not differentially expressed (or negative). Various statistical methods determine $\alpha$ by considering metrics such as sensitivity, specificity, false discovery rate, etc. We compare four methods: raw p-value (i.e. $\alpha=0.05$), Bonferroni multiple test corrected p-value, Benjamini-Hochberg false discovery rate with Storey null hypothesis proportion correction, and our own local false discovery rate estimator.

**Small Physiologically Relevant Changes**

We model small physiologically relevant by using the differential expression density model described in the Simulated Data section with a threshold of 25% and maximum differential expression of 50%. We wish to detect genes that are differentially expressed 25% or more (without penalty of detecting genes with lower differential expression).

Figure 3a shows the number of differentially expressed genes and not differentially expressed genes (as taken from the truth model) of genes ordered by p-value (for 8 observations) and binned in sets of 50. It is clear that, given enough observations (4 or 5 in this example, see supplementary data), smaller p-values yield low probability of being a false positive in a local window. This observation motivates the need for a local *fdr* metric. We estimate this false discovery rate in a local window and draw a threshold when more than 20% (or some other value) of the genes will yield false positives.

The *fdr* and FDR can be estimated from the p-values as described above. Figure 3b shows the *fdr* and FDR metric estimates versus the number of genes classified as differentially expressed for 8 observations. Comparing this to figure 3a, we see that the estimate of *fdr* is close to the actual *fdr* (derived from the truth model). Clearly, the slope of the *fdr* is much higher than the slope of the FDR, and therefore, many different values of *fdr* yield the same number of rejected hypotheses. This makes *fdr* more robust than FDR. Because *fdr* and FDR are not necessarily monotonic functions, it is more appropriate to measure robustness by examining the number of genes classified as differentially expressed as a function of the metric threshold value as shown in figure 3c. The *fdr* method is robust to the threshold parameter value. An *fdr* threshold of 0.1 or 0.2 or 0.3 yields approximately the same number positives. However, FDR thresholds of 0.1, 0.2, or 0.3 yield very different numbers of positives.

Figure 3d illustrates the true and false positives with p-value thresholds provided by several statistical methods for 8 observations. The *fdr* method chooses a p-value threshold where the number of false positives in a small window is a certain percentage of positives in that window. This, in effect finds a threshold where the true positive line begins to deviate from the empirical line. In this example, an *fdr* threshold of 0.2 would classify 254 genes as differentially expressed, which corresponds to FDR of 6%. An *fdr* threshold of 0.5 would classify 331 genes as differentially expressed, which corresponds to FDR of 15%. The FDR method chooses a p-value threshold where the number of false positives is a certain percentage of positives (20% in this example).

Figure 3e shows the fraction of truly differentially expressed genes (above 30% fold change) that were found (sensitivity or truth discovery rate; TDR). Figure 3f shows the fraction of genes that were incorrectly classified as differentially expressed (false discovery rate). Naïve p-value thresholding at 0.05 obtains an unacceptable number of false positives (FDR=53%). Bonferronni correction has the opposite effect of guaranteeing extremely low false positive rates at the expense of classifying very few genes correctly as differentially expressed. False discovery rate method correctly controls the FDR at 20%, providing better results than Bonferronni. The local *fdr* has much lower false discovery rates at the expense of slightly lower numbers of true positives. Figures for other sample sizes are presented in the supplementary data (simulated.low.pdf).

Figure 4 shows the TDR and FDR in the window of genes between the *fdr* threshold (*fdr*=20%) and the FDR threshold (FDR=20%). Note that the *fdr* is controlling for 20% local false discovery rate and therefore the FDR in this window of genes is always above 20%.

**Larger Changes in Differential Expression**

Larger changes between samples can occur in situations such as disease states (i.e. cancer) (Golub, et al., 1999), or different tissue sources (i.e. brain vs. liver). In such cases, fewer observations are necessary to detect most of the changes. We model this situation by using the differential expression density model described in the Simulated

Data section with a uniform distribution cutoff of 100% and maximum differential expression of 800%. We wish to detect genes that are differentially expressed 100% or more (without penalty of detecting genes with lower differential expression).

The simulations above are repeated for this new density function and the results are outlined in Supplemental Figure S1. Figure S1a shows that at 8 observations, smaller p-values yield low probability of false positives in a local window, even more so than in the previously shown simulated data set. The *fdr* estimator (Figure S1b) estimates the true *fdr* (Figure S1a) well. The number of genes classified as differentially expressed is more robust to the *fdr* metric than the FDR metric (Figure S1c). An *fdr* threshold of 0.2 yields 439 differentially expressed genes (FDR=2.5%) while an *fdr* threshold of 0.5 yields 470 differentially expressed genes (FDR=6%). Note how a large change in *fdr* threshold has little effect on the number of genes classified as differentially expressed. The grey region in figure S1d illustrates the region of p-value thresholds and corresponding differentially expressed genes for this wide range of *fdr* thresholds. Again, any *fdr* threshold in the range of 0.2 to 0.5 yields much lower false positive rates at the expense of slightly lower true positive rates compared to FDR=20%. As the number of experiments increase, this becomes more evident. In this case, with 5 observations, one could extract more than 99% of the differentially expressed genes with a very low FDR by using the *fdr* threshold (0.2-0.5). Figures for other sample sizes are presented in the supplementary data (simulated.high.pdf).

Supplemental Figure S2 shows TDR and FDR in the window of genes between the *fdr* threshold (*fdr*=20%) and the FDR threshold (FDR=20%). Again, note that in this region the proportioin of false positives is much higher than 20% as controlled by the *fdr*.

**Case Study: Cancer dataset**

In 1999, (Golub, et al., 1999), published a microarray study of 38 patients, 27 with acute myeloid leukemia (AML) and 11 with acute lymphoblastic leukemia (ALL) with the goal of finding genes that could classify a patient with one of these diseases (data available at http://www.genome.wi.mit.edu/MPR). Later this dataset was used as a case study for comparing of multiple test correction methods (Ge, et al., 2003). We will use this dataset to compare FDR and *fdr* methods.

We follow their normalization and permutation based p-value generation methods. First, the data was filtered by excluding (1) all data less than 100 and more than 16000 and (2) across samples, genes having (max/min)≤5 and (max-min)≤500. Filtering removed 4206 of the 7130 genes leaving 2924 genes. Second, the data was logarithmically transformed (base 10). The p-values of each gene were calculated (following Box 1 of Ge et al, 2003) by permuting the conditions 1,000,000 times and calculating the t values from a two sided T-test for each permutation. Then the t value of the conditions in the correct order is and the percent of t values larger than this value is the p-value.

Figure 5a shows the *fdr*, FDR of the Golub dataset. Additionally, we show the *fdr* as estimated by the twilight package in BioConductor (Scheid and Spang, 2004; Scheid and Spang, 2005). Notice that because the twilight estimator smooths the results, it is overly conservative for small p-values. In all other cases, it trends with our *fdr* estimator. The FDR is fairly smooth, but the *fdr* shows a large jump between 0.8 and 0.14 and then continues to increase. If we use *fdr* thresholds of 0.8, 0.1 and 0.14, we obtain 653, 665, and 692 genes respectively and p-value thresholds of 0.017, 0.019, and 0.022 respectively. This corresponds to a false discovery rate (FDR) of 4%. Assuming this false discovery rate is tolerable, classifying genes with p-values higher than 0.014 as differentially expressed will yield 15% or more additional false positives.

Figure 5b shows the number of genes classified as differentially expressed for all *fdr* thresholds. Typically one would choose an *fdr* threshold that was near the suitable tolerance but in a flat region of Figure 5b. This way, all thresholds in this plateau would yield similar numbers of genes classified as differentially expressed. Note that there are two major plateaus at 0.08<*fdr*<0.14 and 0.16<*fdr*<0.25. Depending on FDR tolerance (4% and 8% respectively), either of these thresholds will yield robust results (653-692 genes and 896-936 genes). Note that the twilight *fdr* estimation has no corresponding plateaus and therefore offers no robust threshold choices.

**DISCUSSION**

We presented a novel way to approach local false discovery rate control based multiple test corrections. We have introduced a new estimator of the local false discovery rate in a window of genes sorted by p-value. This allows researchers to assess the opportunity cost of validating more genes given that they have already validated a set of genes. We have demonstrated the approach on simulated and real gene expression datasets and have presented an algorithm for choosing a p-value threshold given an fdr threshold.

Most local *fdr* estimators are either smoothed or are derived from a parametric equation (such as the beta distribution) resulting in a smooth estimator. In contrast, our proposed method provides finer resolution based on a non-smoothed fluctuating *fdr* estimate within a window of sorted p-values and our heuristics provide robust thresholds. We employ a heuristic approach for the following reasons. First, there are no 'adjusted p-values' as proposed in (Benjamini and Hochberg, 1995) and (Storey, 2003). Choosing the appropriate threshold requires manual assessment of the data and the FDR is still necessary to assess the total amount of Type I error is acceptable. In addition, the *fdr* requires a window size parameter. Smaller window sizes pick up changes faster, but tend to increase noise. Larger window sizes smooth the metric at the cost of robustness. We have presented the *fdr* method with a window size of 50, which we have empirically found to smooth the data sufficiently and yet capture quick changes. For example, if all of the differentially expressed genes have p-values less than those of the genes not changing (as in the simulated bimodal dataset with 7 or more observations), then an *fdr* of 0.2 will only pick 10 false positives (e.g. 10/50 = 0.2).

Whenever discussing *fdr* metric estimates, we have given the corresponding FDR estimate. Because the FDR estimate was near 0.05, one may ask why not use that criteria instead. First, FDR=5% occurred in the simulated data as an artifact of how the problem was setup (i.e. 500 genes differentially, window size of 50, good discrimination of differentially expressed genes, etc). It is entirely possible that in another scenario, the FDR would be 0.005 (e.g. 5000 genes differentially expressed instead of 500). 5% is an arbitrary threshold. Further, in the Golub et al dataset, *fdr* began to sharply rise when FDR was 4%. However, the *fdr* provides a similar FDR threshold for most ranges of the metric (e.g. 0.2-0.3). Based on the results presented here, for a given global FDR tolerance interval, our local *fdr* estimator provides a systematic way to obtain robust thresholds for multiple hypotheses testing.

**GRANTS**

This work was supported by NIH/NIAAA award R01 AA13204, DARPA award F30602-01-2-0578, and NIGMS award R01 P20 GM67266 to JSS and by NIAAA training grant support and a Greater Philadelphia Bioinformatics Alliance Fellowship award to RLK.

**FIGURES**

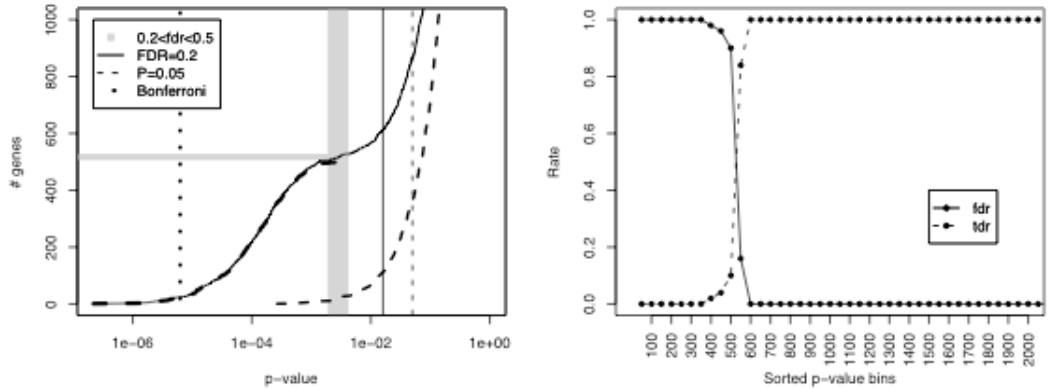

Figure 1: (a) P-values vs number of genes classified as differentially expressed (rejected hypotheses). Solid curve represents the empirical p-values. Thick dashed curve represents the p-values of the differentially expressed genes (as determined by the truth model). Thin dashed curve represents the p-values of the genes not differentially expressed (as determined by the truth model). The legend corresponds to the vertical lines representing various thresholds. (b) P-values are ranked from smallest to largest and binned into groups of 50. *fdr* and *tdr* are the proportion of false and true positives, respectively, as determined from the truth model in each bin. Note that after 500 genes, the *fdr* within the bin becomes very high. However, the FDR of all 500 genes is still low (1.6%). This is the motivation for using local false discovery rate.

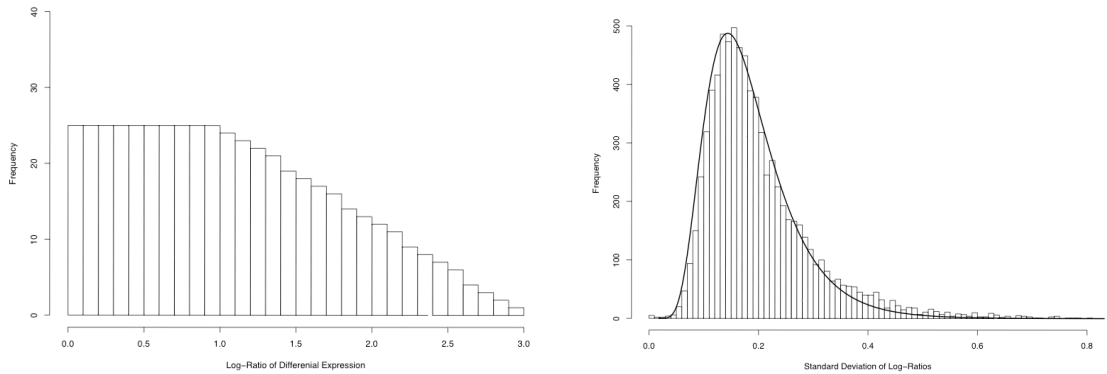

Figure 2: Parameterization of simulated differential expression and noise model. (a) Distribution of differentially expressed genes, uniform to a cutoff and tapered to a maximum expression. In this example the cutoff is 1.0 and the maximum expression is 3.0. (b) Histogram of the standard deviation of log-ratio differential gene expression from a microarray experiment with 6 pairs of biological replicates. The solid line is the log-normal distribution with $\mu$=-2.8 and $\sigma$=0.6.

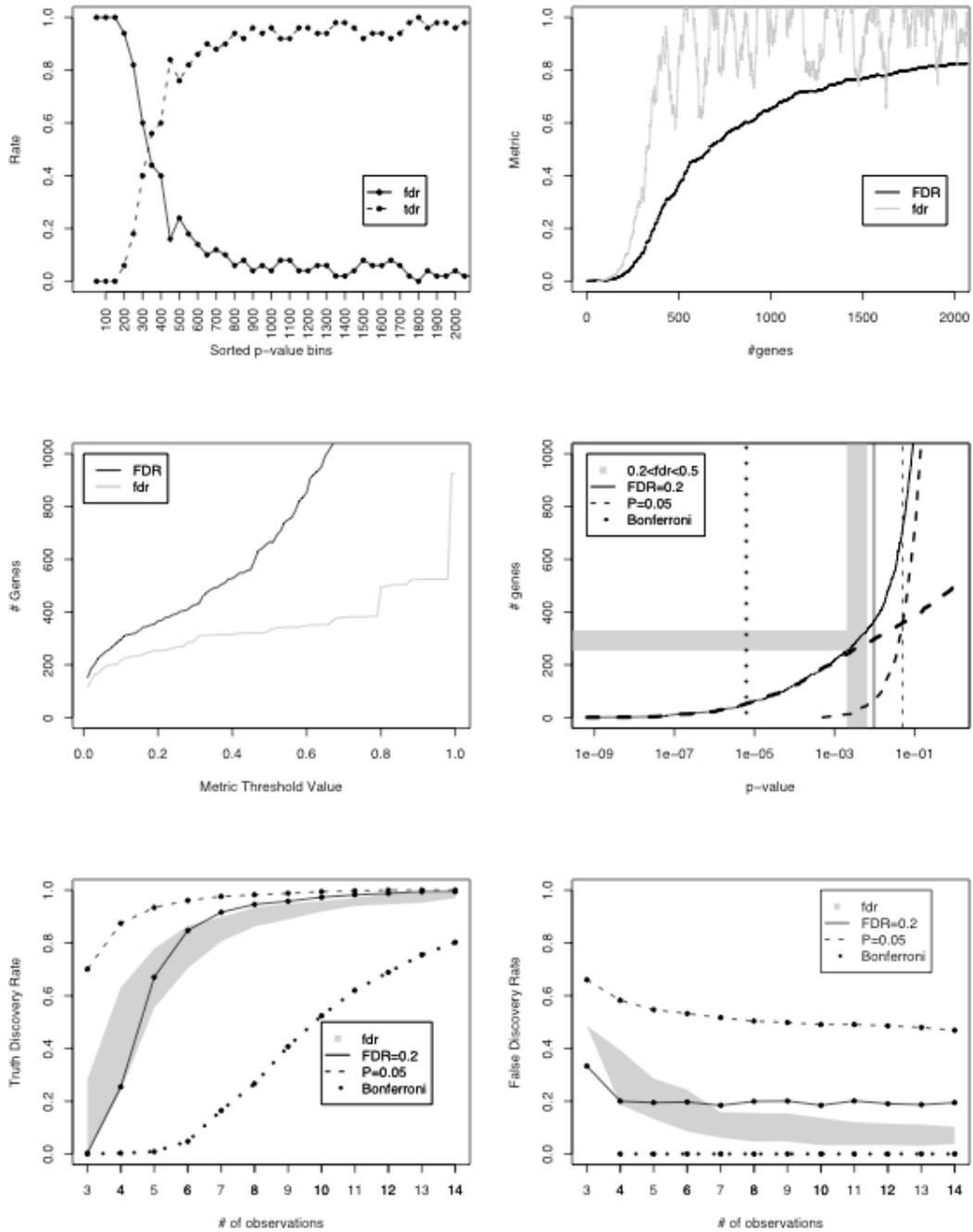

Figure 3: Small changes case study. (a) P-values are ranked from smallest to largest and binned into groups of 50. *fdr* and *tdr* are the proportion of false and true positives, respectively, as determined from the truth model in each bin. (b) FDR and local *fdr*

estimates. Note that the local *fdr* is a good estimate of the true local *fdr* as shown in Figure 3a. (c) Number of genes classified as differentially expressed at various metric threshold values for FDR and local *fdr*. Note that the local *fdr* forms many plateaus where very different local *fdr* threshold values yield similar numbers of rejected hypotheses, whereas the FDR does not have this characteristic. (d) P-values vs number of genes classified as differentially expressed (rejected hypotheses). Solid curve represents the empirical p-values. Thick dashed curve represents the p-values of the differentially expressed genes (as determined by the truth model). Thin dashed curve represents the p-values of the genes not differentially expressed (as determined by the truth model). The legend corresponds to the vertical lines representing various thresholds. Number of differentially expressed genes: 254-331 for 0.2<*fdr*<0.5, 366 for FDR<0.2 (e) Sensitivity based on the truth model vs sample size. Each point represents the median of 10 runs. The grey region represents the region between 0.2<*fdr*<0.5. Note that the FDR and the local *fdr* thresholds yield similar sensitivities. Also, after 6 observations, over 90% of all differentially expressed genes are correctly identified. (f) Actual FDR based on the truth model vs sample size. Each point represents the median of 10 runs. The local *fdr* has much lower false discovery rates at the expense of slightly lower numbers of true positives.

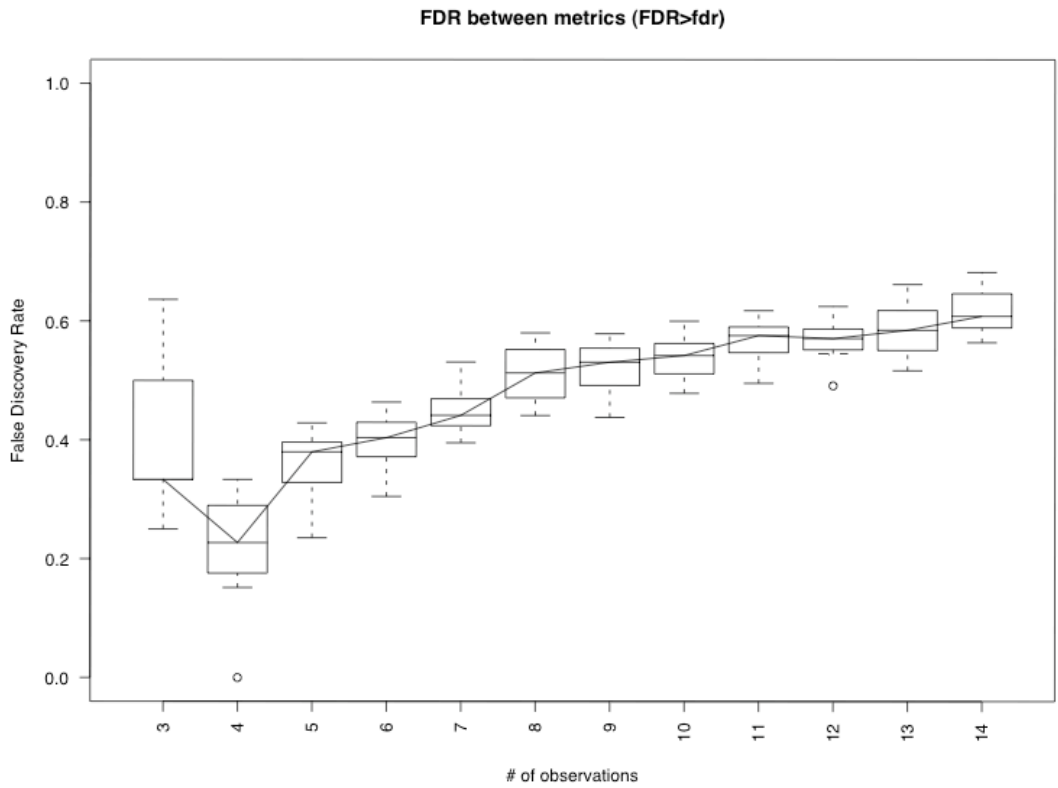

Figure 4: Small changes case study. Boxplot of the additional false positive proportion incurred by choosing FDR<0.2 instead of *fdr*<0.2. Note that after five samples, all runs result in an additional false positive proportion higher than the FDR threshold (0.2).

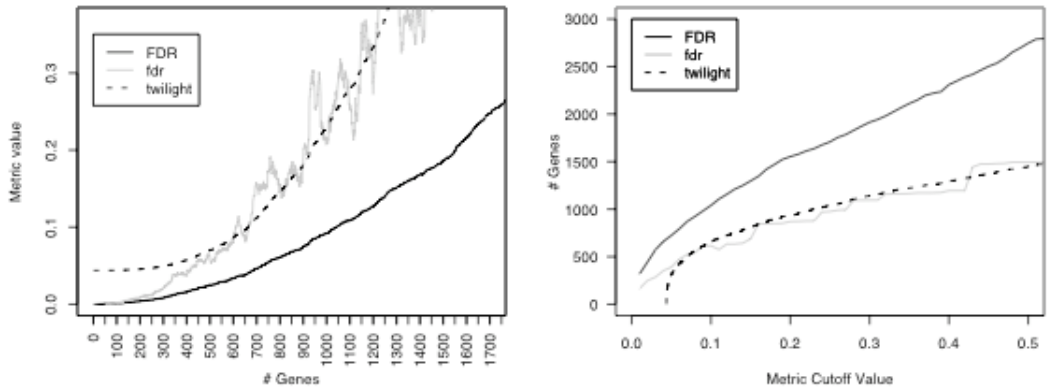

Figure 5: Cancer dataset. (a) FDR (solid line), our local *fdr* estimate (grey solid line), and twilight local *fdr* estimate (dashed line). Note that the twilight estimate is smoothed and is conservative for small p-values (i.e. first 500 genes). For large p-values the twilight estimate follows the trend of our estimate. (b) Number of genes classified as differentially expressed (rejected hypotheses) at various threshold values for FDR (solid line), our local *fdr* estimator (grey line), and the twilight local *fdr* estimator (dashed line). Our local *fdr* estimator provides a number of plateaus including two at 0.08<*fdr*<0.14 and 0.16<*fdr*<0.25. The heuristics to determine a robust threshold choice are discussed in the text. The twilight local *fdr* estimator does not provide any clear robust thresholds.

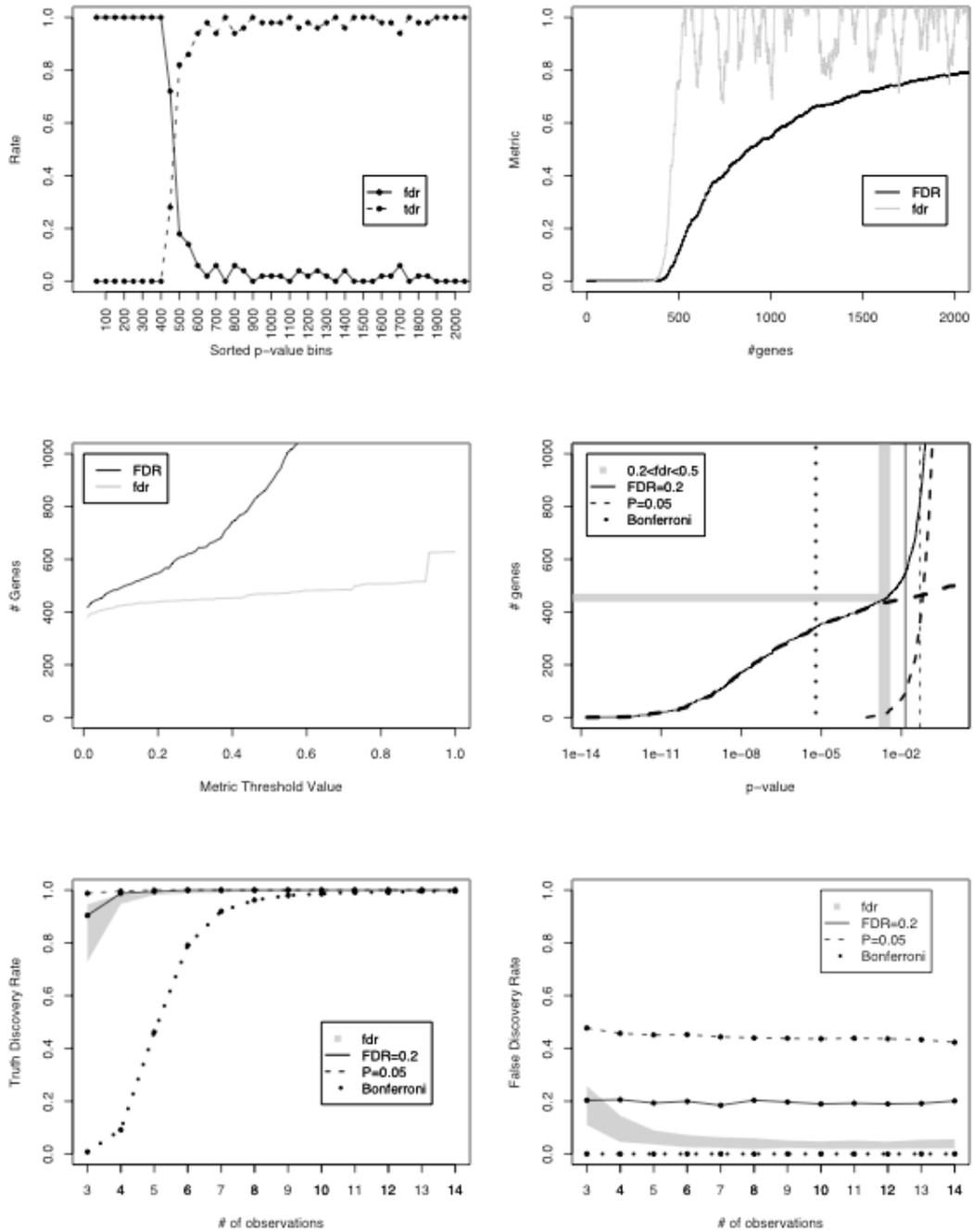

Figure S1: Large changes case study. Same plots as Figure 3 for the simulated data with large changes. In (d), the number of differentially expressed genes are: 439-470 for 0.2<*fdr*<0.5, 547 for FDR<0.2.

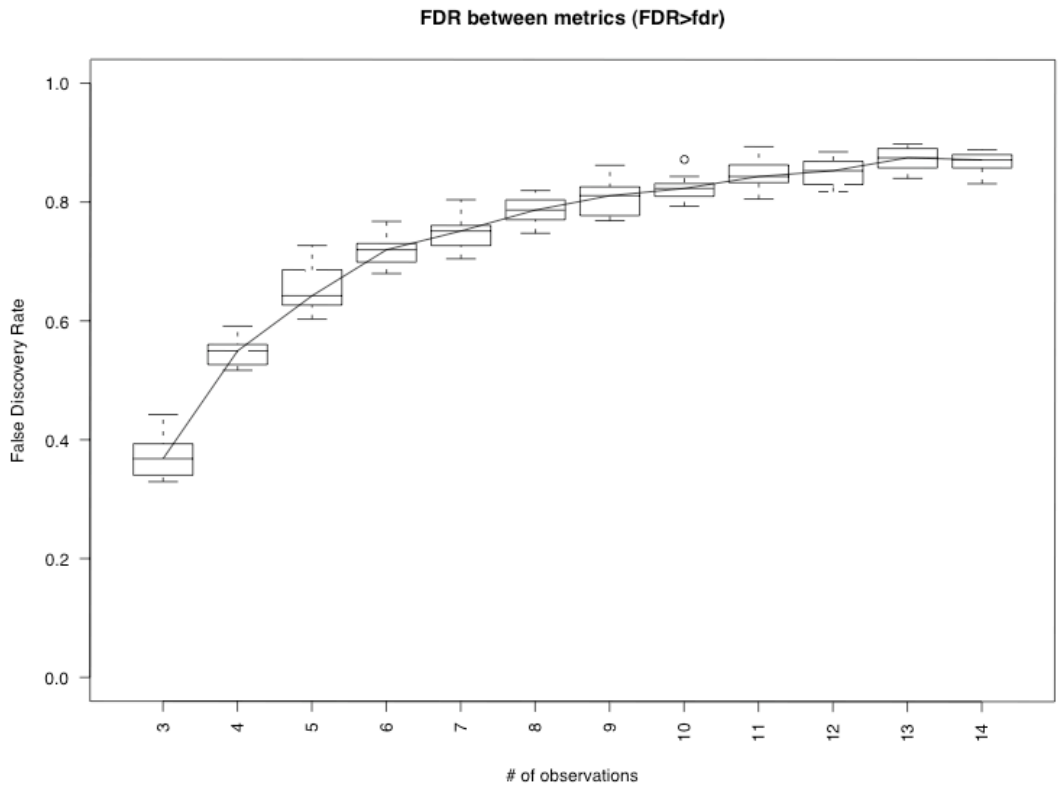

Figure S2: Large changes case study. Boxplot of the additional false positive proportion incurred by choosing FDR<0.2 instead of *fdr*<0.2. Note that for all sample sizes, all runs result in an additional false positive proportion higher than the FDR threshold (0.2).